\documentclass{aa}

\usepackage{psfig}
\usepackage{graphicx}

\def\etal{{et\,al.}}
\def\asec{\ifmmode ^{\prime\prime}\else$^{\prime\prime}$\fi}
\def\farcs{\hbox{$.\!\!^{\prime\prime}$}}  
\def\msun{$M_{\odot}$}
\newbox\grsign \setbox\grsign=\hbox{$>$}
\newdimen\grdimen \grdimen=\ht\grsign
\newbox\laxbox \newbox\gaxbox
\setbox\gaxbox=\hbox{\raise.5ex\hbox{$>$}\llap
     {\lower.5ex\hbox{$\sim$}}}\ht1=\grdimen\dp1=0pt
\setbox\laxbox=\hbox{\raise.5ex\hbox{$<$}\llap
     {\lower.5ex\hbox{$\sim$}}}\ht2=\grdimen\dp2=0pt

\def\lax{$\mathrel{\copy\laxbox}$}

\begin{document}

\title{Identification of the donor in the X-ray binary 
GRS 1915+105\thanks{Based on 
observations collected at the European Southern Obser\-vatory,
Chile under proposals ESO Nos. 63.H-0465 and 65.H-0422.}} 

\authorrunning{Greiner, Cuby, McCaughrean et al.}
\titlerunning{The donor in GRS 1915+105}

\author{J. Greiner\inst{1}, J.G. Cuby\inst{2}, M.J. McCaughrean\inst{1}, 
A.J. Castro-Tirado\inst{3,4}, R.E. Mennickent\inst{5}}

\offprints{J. Greiner, jgreiner@aip.de}

\institute{Astrophysikalisches Institut Potsdam, 14482 Potsdam, 
      An der Sternwarte 16, Germany
   \and
    ESO, Alonso de C\'ordova 3107, Santiago 19, Chile
   \and
    Instituto de Astrof\'{\i}sica de Andaluc\'{\i}a, (IAA-CSIC), 
     P.O. Box 03004, E-18080 Granada, Spain
   \and
   Laboratorio de Astrof\'{\i}sica Espacial y F\'{\i}sica Fundamental
   (LAEFF-INTA), P.O. Box 50727, E-208080 Madrid, Spain
   \and
   Universidad de Concepci\'on, Casilla 160-C Concepci\'on, Chile}

\date{Received 14 May 2001; accepted 26 May 2001}

\abstract{
We report on the results of medium-resolution spectroscopy of GRS 1915+105
in the near-infrared H and K band using the 8m VLT at ESO.
We clearly identify absorption bandheads from $^{12}$CO and $^{13}$CO.
Together with other features this results in a classification of the 
mass-donating star in this binary as a K-M\,III star, clearly indicating 
that GRS 1915+105
belongs to the class of low-mass X-ray binaries (LMXB).
\keywords{stars: binaries -- infrared: stars -- 
     stars: individual: GRS 1915+105}
}

\maketitle

\section{Introduction}  

GRS 1915+105 (Castro-Tirado \etal\ 1994) is the prototypical microquasar, 
a galactic X-ray binary
ejecting plasma clouds at v$\approx$0.92\,c (Mirabel \& Rodriguez 1994).
It exhibits unique X-ray variability patterns (Greiner \etal\ 1996) 
which have been interpreted as
accretion disk instabilities leading to an infall of parts of the
inner accretion disk (Belloni \etal\ 1997).
Based on its X-ray properties GRS 1915+105 is suspected to be the 
most massive stellar black hole candidate in
the Galaxy (Morgan \etal\ 1997). It is one of only 
two galactic sources which are thought
to contain a maximally spinning black hole (Zhang \etal\ 1997).
It is therefore of great importance to know some details about the
system components in order to understand the conditions which lead to
the unique X-ray, radio, and IR characteristics.

Previous infrared spectroscopy at Palomar 
(Eikenberry \etal\ 1998), UKIRT (Mirabel \etal\ 1997, Harlaftis \etal\ 2001)
and the VLT (Mart\'{\i} \etal\ 2000)
has shown that the \ion{He}{II} emission line is variable, most probably 
depending on the X-ray activity. In particular, Eikenberry \etal\ (1998)
found variations in line flux of 5 on 5--10 min. timescales, 
suggesting that these IR lines are radiatively pumped by (presumably) 
jet ejection events rather than high X-ray luminosity. 
Based on the IR spectral variability and on its
position on the IR H-R diagram, Castro-Tirado \etal\ (1996) suggested
that GRS 1915+105 is a LMXB. On the contrary, 
based on the detection of \ion{He}{I}, Mirabel \etal\ (1997)
and Mart\'{\i} \etal\ (2000) suggested that the donor in GRS 1915+105
is a high-mass O or B star, and that accretion occurs predominantly
from the wind of the donor.

\section{Observations and Results}

We wished to search for absorption signatures due to the donor
in the binary system GRS 1915+105. Since
GRS 1915+105 is a strongly variable infrared source, believed to 
predominantly caused by synchrotron emission of ejected material
(Eikenberry \etal\ 2000, Greiner \etal\ 2001), this required high 
signal-to-noise as well as good spectral resolution in order to beat the 
strong veiling. We therefore used the infrared spectrometer ISAAC 
on the 8m VLT Antu telescope on Paranal (ESO, Chile) to obtain H and K band
spectra.

The short wavelength (0.9--2.5 $\mu$m) arm of ISAAC is
equipped with a 1024$\times$1024 pixel Rockwell HgCdTe array
with an image scale of 0\farcs147/pixel.
Using the medium
resolution grating (0.8 \AA/pixel in the H band, 1.2 \AA/pixel in the K band)
yields a spectral resolution of $\sim$3000 with a 1\asec\ slit.
Observations were performed in 1 (2) adjacent H bands, and 2 (3) 
adjacent K bands on 20/21 July 1999 (24/25 July 2000).

Science exposures consisted of several 250--300 sec individual exposures
which were dithered along the slit by $\pm$30\asec.
In order to correct for atmospheric absorption, the nearby star HD 179913
(A0\,V) was observed either before or after each science exposure.
The initial data reduction steps like bias subtraction,
flatfielding and co-adding
were performed within the {\em Eclipse} package (Devillard 2000).
The extraction and wavelength calibration was done using an optimal extraction
routine within the MIDAS package.

 \begin{figure*}[th]
   \centering{\vspace{-0.2cm}
   \hspace{-0.01cm}
   \vbox{\psfig{figure=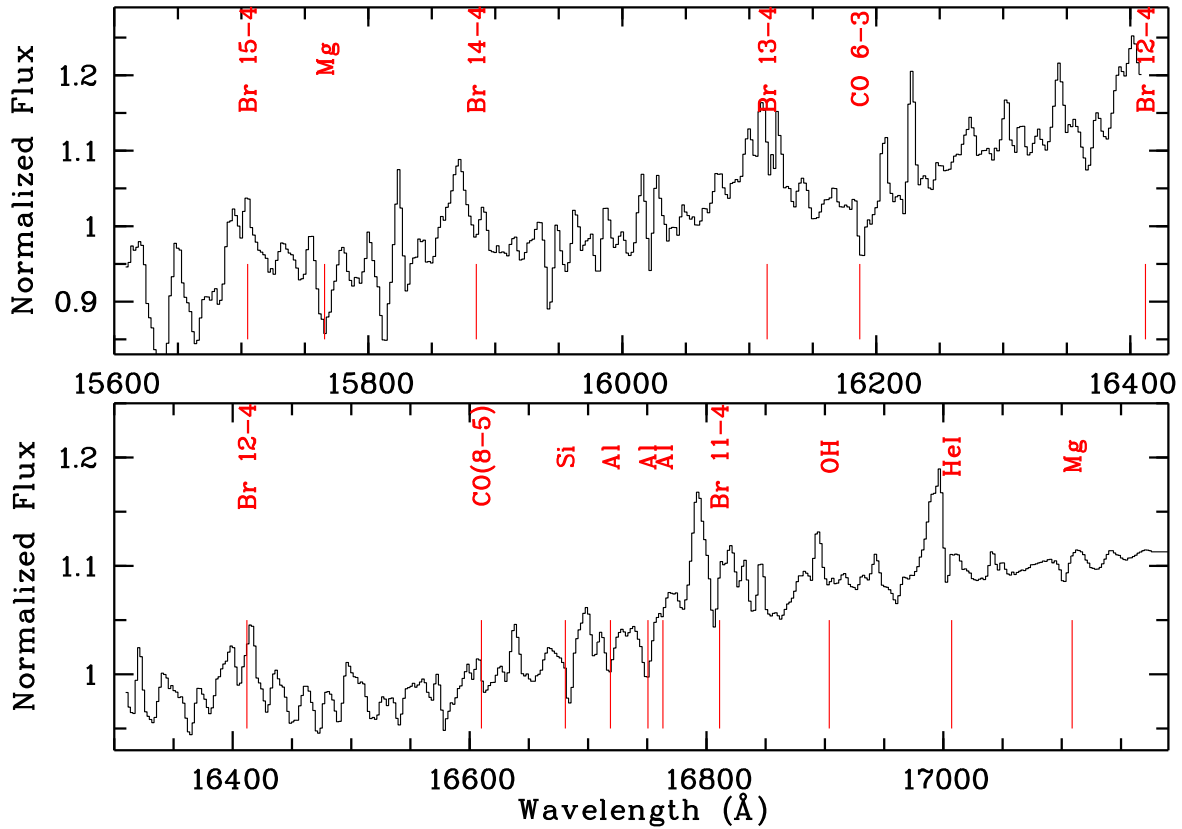,width=0.87\textwidth,%
         bbllx=2.2cm,bblly=1.2cm,bburx=19.3cm,bbury=13.8cm,clip=}}\par
   \vspace{-0.2cm}
   \caption[spec]{Normalized spectrum of GRS 1915+105 in the H band.
          The top panel represents a 6000 sec exposure from 1999, 
          while the bottom panel shows 
          the sum of the 2 images taken in 1999 and 2000
          (total exposure of 8400 sec).
          The spectra were rebinned to 2 \AA\ resolution.
          Obvious detected lines are marked by vertical lines and labels,
          as well as the location of lines mentioned in the text.
         \label{hspec}}}
\end{figure*}

\begin{table}
\caption{Observation log}
\vspace{-0.3cm}
\begin{tabular}{ccccc}
      \hline
      \noalign{\smallskip}
       Wavelength & $\!\!\!$21-07-1999$\!\!\!$ & $\!\!\!$Exposure & 
                    $\!\!\!$24-07-2000$\!\!\!$ & $\!\!\!$Exposure$\!\!$  \\
      \noalign{\smallskip}
      \hline
      \noalign{\smallskip}
  $\!\!$1.56--1.64 $\mu$m & & & 0:33--2:15$\!\!$ & $\!\!$10*600 sec$\!\!\!$ \\
  $\!\!$1.63--1.72 $\mu$m &     5:43--6:25$\!\!$ & $\!\!$\,\,8*300 sec$\!\!\!$ 
                             &  2:20--4:03$\!\!$ & $\!\!$10*600 sec$\!\!\!$ \\
  $\!\!$2.06--2.17 $\mu$m &     4:20--4:59$\!\!$ & $\!\!$12*180 sec$\!\!\!$ 
                             &  4:19--5:10$\!\!$ & $\!\!$10*300 sec$\!\!\!$ \\
  $\!\!$2.17--2.29 $\mu$m & & & 5:11--6:03$\!\!$ & $\!\!$10*300 sec$\!\!\!$ \\
  $\!\!$2.29--2.41 $\mu$m &     6:35--7:18$\!\!$ & $\!\!$\,\,8*300 sec$\!\!\!$
                             &  6:21--7:31$\!\!$ & $\!\!$10*300 sec$\!\!\!$ \\
 \noalign{\smallskip}
 \hline
\end{tabular}
\end{table}

 \begin{figure*}[th]
   \centering{\vspace{-0.2cm}
   \hspace{-0.01cm}
   \vbox{\psfig{figure=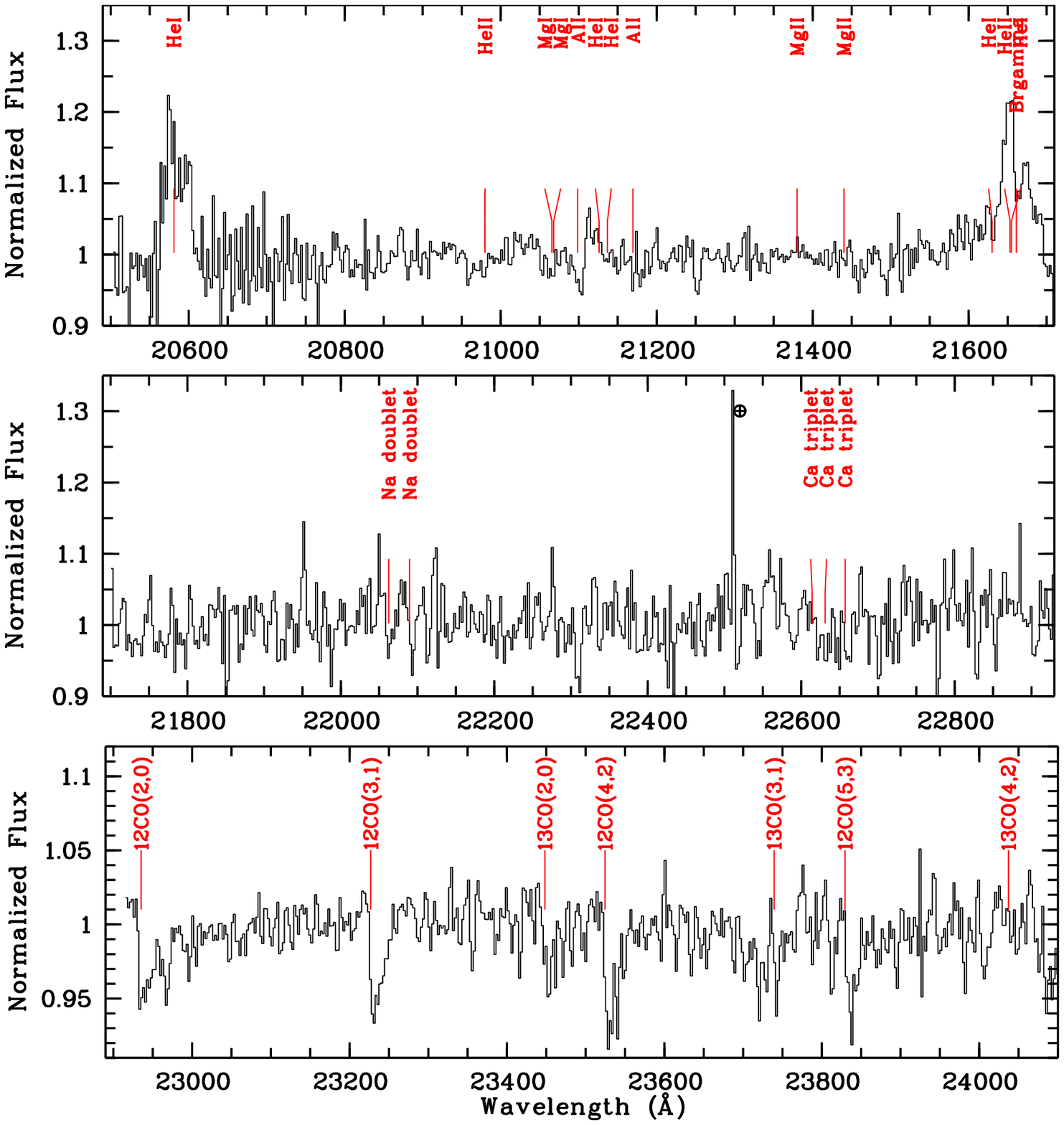,width=0.87\textwidth,%
         bbllx=2.2cm,bblly=7.7cm,bburx=19.4cm,bbury=26.9cm,clip=}}\par
   \vspace{-0.3cm}
   \caption[spec]{Spectrum of GRS 1915+105 in the K band, normalized by
          division through the continuum as derived from a spline fit. 
          The top two panels
          correspond to 3000 sec exposure, while the bottom panel shows 
          a sum of 4 images (total exposure 9400 sec). All spectra were 
          rebinned by a factor of 2.
         \label{kspec}}}
\end{figure*}

The spectra of the 
2 grating settings covering the H band are shown in Fig. \ref{hspec}, and 
of the 3 grating settings covering the K band in Fig. \ref{kspec}.
The K band top panel  shows the strong Br\,$\gamma$ and He\,I
emission lines known from previous low-resolution spectroscopy
(Castro-Tirado \etal\ 1996). Similarly, we see Brackett series in emission
from 11-4 (Br\,$\eta$) to 15-4 (Br\,$\lambda$) in the H band.
We do not find \ion{He}{II} emission as reported by Castro-Tirado \etal\
(1996), Eikenberry \etal\ (1998) and Mart\'{\i} \etal\ (2000), supporting 
their conclusion that this is
a variable feature probably related to the X-ray state and jet ejection 
activity. We note that during our 1999 ISAAC observation of GRS 1915+105 
was in a state of low activity at X-rays and radio, though the time from 
the last radio flare and towards the next radio flare were different to the
Mart\'{\i} \etal\ (2000) observation.

Both, the H as well as \ion{He}{I} lines are clearly resolved, having 
FWHM $\sim$ 10--15 \AA\ at a resolution of 5 \AA\ (in the H band). 
If this were due to rotational Doppler broadening, it would correspond to a 
velocity of $v$ sin$i$ $\sim$ 200--300 km/s.
In most cases, these H/\ion{He}{I} lines are not gaussian, but have a central
depression. Given the fact that the inclination of the binary system is 
$i \sim 70\deg\pm2\deg$ (Mirabel \& Rodriguez 1994) one indeed may anticipate 
a double-lined shape. 
We do not find P Cyg profiles in the Br\,$\gamma$ and He\,I as reported
by Mart\'{\i} \etal\ (2000).
 
In addition, we find
for the first time several absorption lines which allow us to make a rough
identification of the donor in the GRS 1915+105 binary.
In the K band we clearly identify $^{12}$CO absorption band heads 
characteristic 
of a low temperature ($T< 7000$ K) star (e.g. Kleinmann \& Hall 1986). 
Though weak, we also identify the 
$^{13}$CO (2,0) and $^{13}$CO (3,1) transitions, indicating a luminosity
class III or brighter (e.g. Wallace \& Hinkle 1997). 
We also identify the Na doublet (2.20624/2.20897 $\mu$m), and possibly
the Ca triplet (2.26141/2.26311/2.26573\,$\mu$m), Al\,I (2.10988 $\mu$m)
and the Mg\,I doublet (2.10655/2.10680\,$\mu$m) in absorption.
Note that the CN doublet (2.0910/2.0960\,$\mu$m), which in supergiants
is more prominent than Al/Mg, is not detected. 
In the H band, we identify Mg\,I (1.5749\,$\mu$m) (though $^{12}$CO (4,1)
may also contribute), $^{12}$CO (6,3) and $^{12}$CO (8,5) in a ratio which 
is consistent with MK standards (Meyer \etal\ 1998), 
and Al\,I (16718.9/16750.6 $\mu$m).
Comparing the 2.3--2.4 $mu$m spectrum from 20/21 July 1999 with that 
taken on 24/25 July 2000 (after heliocentric correction), we find that 
the CO band head systems are shifted by 60 km/s relativ to each other.
The easiest interpretation is Doppler motion, and therefore indicates that 
the CO absorption is indeed of photospheric origin and not due to absorption
in a static, cold, circumstellar medium.
Thus, we conclude that we have identified the donor in GRS 1915+105, and 
that it is a late-type, K-M giant.

We have tried to confirm the luminosity class more quantitatively
by using the veiling-independent indicator \\
 $~~~~~r = \log [EW(^{12}{\rm CO} (2,0))/(EW({\rm Na}) + EW({\rm Ca}))]$ \\
(Ram\'{\i}rez \etal\ 1997).
Because of the low significance of the Ca triplet, our measurement
has a large error: $r = 0.25 \pm 0.20$. This value falls in between
the ranges covered by dwarfs (--0.2\lax $r$ \lax 0.0) and
giants (0.4\lax $r$ \lax 0.6) (Ram\'{\i}rez \etal\ 1997).
The ratio of equivalent widths of $^{12}$CO to $^{13}$CO 
which depends on luminosity class (Campbell \etal\ 1990), has been measured
for the seven transitions covered (lower panel of Fig. \ref{kspec})
to $\sim 3 \pm 1$, again supporting a giant classification.

 \begin{figure*}[th]
  \sidecaption
  \vspace{-0.2cm}
  \includegraphics[width=12.7cm]{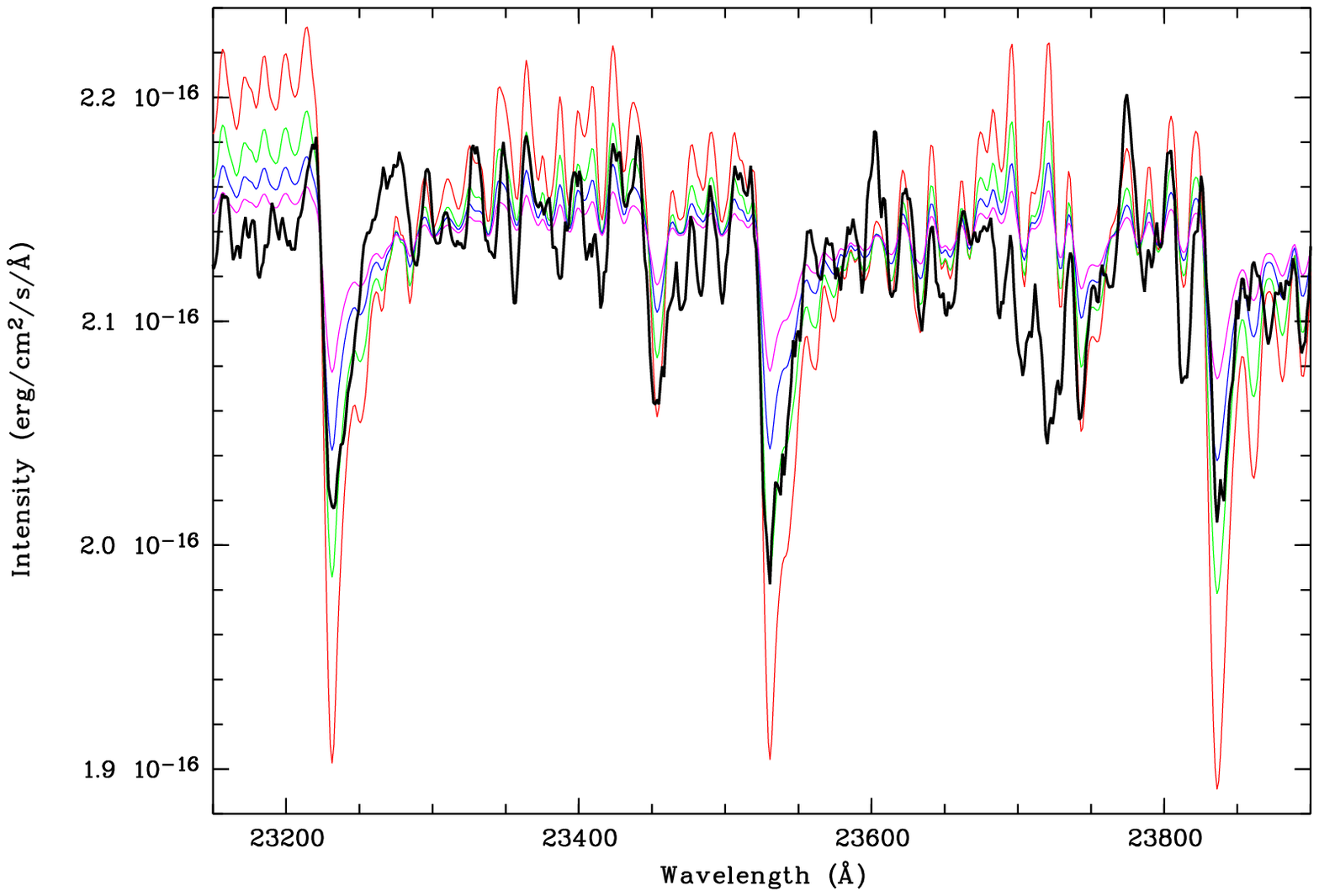}
   \caption[veil]{Spectrum of GRS 1915+105 in the 2.32--2.39 $\mu$m range 
      (thick black line; uncorrected for extinction) 
      compared to the spectrum
      of the K2III star HD 202135 scaled to magnitudes of 14.0 (red), 
      14.5 (green), 15.0 (blue) and 15.5 (pink) and
      artificially veiled with a flat continuum such that the total 
      K band brightness is 13.2 mag. Comparing the equivalent widths
      of the $^{12}$CO(3,1) $^{12}$CO(4,2) and $^{12}$CO(5,3) bandheads
      between the GRS 1915+105 spectrum (-1.3 \AA, -2.3 \AA, -0.7 \AA) 
      and those 4 ``synthetic'' spectra 
      (red: -4.0 \AA, -3.1 \AA, -2.4 \AA; green: -2.6 \AA, -1.9 \AA, -1.5 \AA; 
       blue: -1.5 \AA, -1.2 \AA, -0.9 \AA; pink: -0.8 \AA, -0.7 \AA, -0.7 \AA)
      implies that the donor of GRS 1915+105 has an unveiled
      magnitude of 14.5--15 mag (prior to extinction correction).\\

      \label{veil}}
 \end{figure*}

Using the H band spectra, we also considered the veiling-independent
temperature/luminosity discriminant 
EW(OH 1.6904 $\mu$m)/EW(Mg 1.5765 $\mu$m) vs. 
EW(CO 1.6610 $\mu$m + CO 1.6187 $\mu$m)/EW(Mg 1.5765 $\mu$m)
as proposed by Meyer \etal\ (1998).
The OH line is, unfortunately, only marginally detected, and therefore
only the luminosity class cannot be constrained. The
temperature estimate yields
$\sim$4800$^{+200}_{-500}$ K, which would suggest a late-G or K 
spectral type (Houdashelt \etal\ 2000), but since the CO bands are weak,
the error again is too large to allow more detailed refinements.

In order to determine the luminosity of the donor, the veiling was 
roughly determined by comparison of our flux-calibrated 2.35\,$\mu$m spectrum 
of GRS 1915+105 with that of a K2\,III standard star, observed with the same
settings. The flux calibration of the GRS 1915+105 spectrum was done
using the acquisition image taken immediately before the spectrum
which itself was photometrically calibrated with local field standards
(see Greiner \etal\ 2001) to yield K = 13.2 mag for GRS 1915+105.
We therefore took the K2\,III star
spectrum, scaled to a brightness in the range 13.5--16.0 mag
and added a flat continuum such that the total K band brightness is 13.2 mag.
A comparison of the depths of the CO band heads with those in the  
GRS 1915+105 spectrum (Fig. \ref{veil}) gives a magnitude of 
K = 14.5--15.0 for the donor, uncorrected for extinction.
With a distance of $\sim$11 kpc (Fender \etal\ 1999) and an
$A_{\rm K} = 3$ mag (Greiner \etal\ 1994, Nagase \etal\ 1994, 
Chaty \etal\ 1996)
extinction correction, this implies an absolute magnitude of 
$M_{\rm K}$ = --2...--3, consistent with the giant classification.

This identification of the donor of GRS 1915+105 as a K-M giant
implies a rather narrow mass range of 1.0--1.5 \msun. 
Typical spherical mass-loss rates of such stars are much too low to sustain
the high accretion luminosity of GRS 1915+105 via accretion from 
the donor's stellar wind. We therefore suggest (not too surprisingly)
that accretion should occur via Roche lobe overflow.
This is consistent
with the constraints derived by Eikenberry \& Bandyopadhyay (2000).

We note that our identification of GRS 1915+105 as a LMXB, as earlier 
proposed by Castro-Tirado \etal\ (1996), 
contradicts the findings of Mirabel \etal\ (1997) and
Mart\'{\i} \etal\ (2000) who, on the basis of VLT-ISAAC spectra of lower
resolution and S/N ratio, argued for a massive
OB-type companion. Our donor identification suggests a 
origin of the \ion{He}{I} lines different from the photosphere of the
OB star (Mirabel \etal\ 1997, Mart\'{\i} \etal\ 2000).
In fact, \ion{He}{I} emission is often observed in LMXBs and cataclysmic
variables to be uncorrelated with the donor, but originating presumably
in the accretion disk or disk wind.

The presence of clear donor absorption features will now allow 
a search for the periodic Doppler shifts due to orbital motion,
as well as a determination of the mass of the compact
object using radial velocity measurements of the donor.

\begin{acknowledgements}
We are grateful to the ESO-Paranal staff for executing the July 2000
observations in service mode.
\end{acknowledgements}

\end{document}